\documentclass[twocolumn,showpacs,preprintnumbers,amsmath,amssymb,prl,superscriptaddress]{revtex4-1}
\usepackage{graphicx}
\usepackage{amsmath}
\usepackage{amssymb}
\usepackage{dcolumn}
\usepackage{color}
\usepackage{subeqnarray}
\usepackage[utf8]{inputenc}
\usepackage[english]{babel}
\usepackage{mathrsfs}
\usepackage{wasysym}

\begin{document}

\preprint{}

\title{Hybrid entanglement between optical discrete polarizations and continuous quadrature variables}

\author{Jianming Wen}
\affiliation{Department of Physics, Kennesaw State University, Marietta, Georgia 30060, USA}

\author{Irina Novikova}
\affiliation{Department of Physics, College of William and Mary, Williamsburg, Virginia 23187, USA}

\author{Chen Qian}
\affiliation{Department of Computer Science and Engineering, University of California at Santa Cruz, Santa Cruz, California 95064, USA}

\author{Chuanwei Zhang}
\affiliation{Department of Physics, The University of Texas at Dallas, Richardson, Texas 75080, USA}

\author{Shengwang Du}\email{Corresponding author: dusw@utdallas.edu}
\affiliation{Department of Physics, The University of Texas at Dallas, Richardson, Texas 75080, USA}

\date{\today }

\begin{abstract}
By coherently combining advantages while largely avoiding limitations of two mainstream platforms, optical hybrid entanglement involving both discrete and continuous variables has recently garnered widespread attention and emerged as a promising idea for building heterogenous quantum networks. Different from previous results, here we propose a new scheme to remotely generate hybrid entanglement between discrete-polarization and continuous-quadrature optical qubits heralded by two-photon Bell state measurement. As a novel nonclassical light resource, we further utilize it to discuss two examples of ways -- entanglement swapping and quantum teloportation -- in which quantum information processing and communications could make use of this hybrid technique.
\end{abstract}

\maketitle

\begin{figure*} [t]
\centering
\includegraphics[width=17 cm]{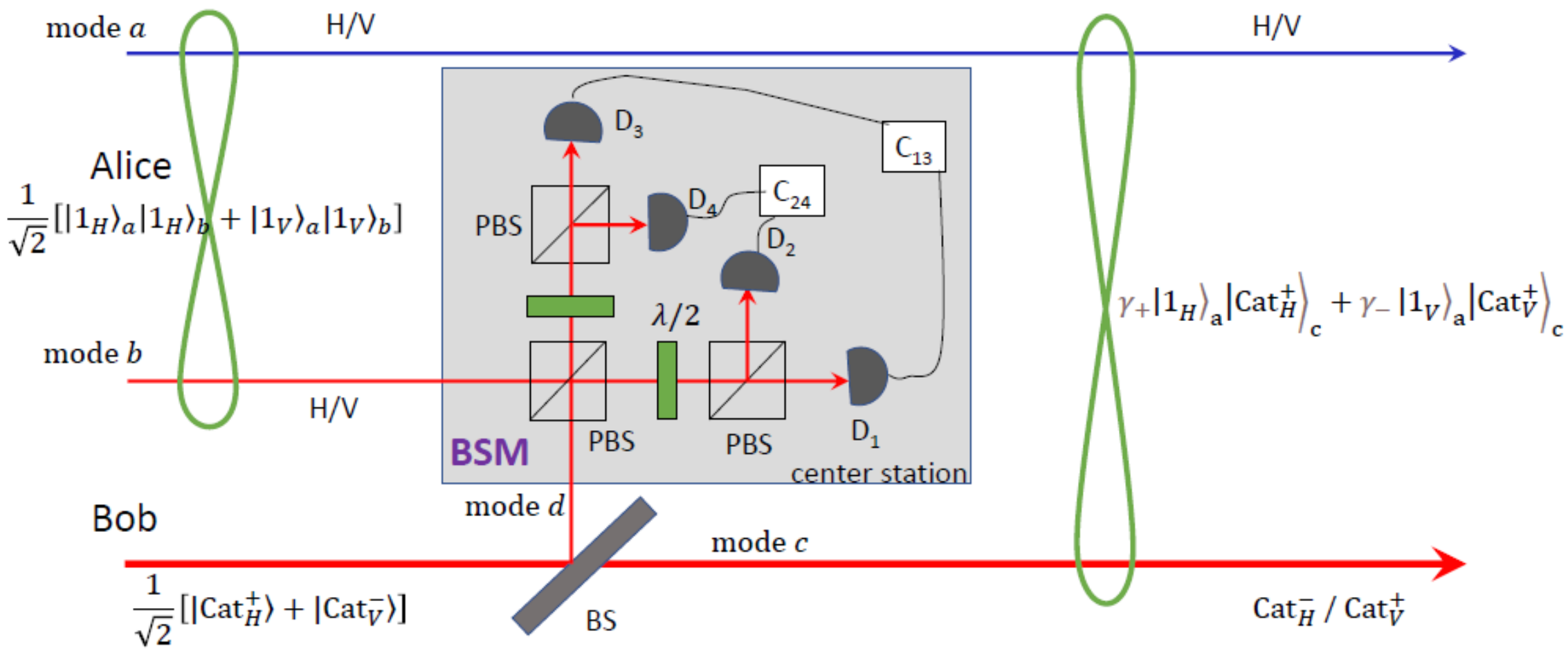}
\caption{(color online) Conceptual scheme for generating hybrid entanglement between single-photon DV polarizations and CV cat states at a distance. Alice sends one mode $b$ of a two-photon polarization entangled state $(|1_H\rangle_a|1_H\rangle_b+|1_V\rangle_a|1_V\rangle_b)/\sqrt{2}$ towards a center station (gray area), while Bob transmits a small portion of a polarization-coupled cat state $(|\mathrm{Cat}^{+}_H\rangle+|\mathrm{Cat}^{-}_V\rangle)/\sqrt{2}$. The two modes are superposed and ready for a two-photon joint Bell state measurement (BSM). Each successful coincidence event between single-photon detectors D$_1$ and D$_3$ (or D$_2$ and D$_4$) heralds the formation of the targeted state (1) between Alice and Bob, which will serve as a powerful resource for further information processing and quantum networks. $\lambda/2$: half-wave plate.}
\label{fig:figure1}
\end{figure*}

Quantum entanglement is an indispensable resource for implementing optical quantum information processing (QIP) [1-7]. Traditionally, two classes of methodologies have been fostered in parallel for entanglement acquisition by exploiting one of the two incompatible facets of light in accordance with the wave-particle duality. Consequently, these developments lead to two distinct directions in encoding information by utilizing either discrete-variable (DV) states of finite dimensions (such as photon number and light polarization) [1-4] or continuous-variable (CV) states of infinite-dimensional Hilbert space  (such as field quadrature components) [5-7]. In practice, both encodings have displayed their respective advantages but also exposed individual weaknesses. With less concern of photon loss, the DV protocols involving single photons generally enjoy almost unit-fidelity, but rely on probabilistic implementations and high-efficient single-photon detectors. In contrast, the CV alternatives, using quadrature components of an electromagnetic field, celebrate unambiguous state discriminations, unconditional operations and perfect homodyne detection efficiency, but suffer from photon loss and inherent low state fidelity because of its coupling to vacuum. Recently, notable efforts [8-22] have been devoted to harnessing merits of both approaches so as to overcome the intrinsic individual limitations. Progress on integrating DV and CV technologies in unified hybrid architectures has manifested the capability of distributing and interconnecting optical DV and CV quantum states (or qubits). One could envision a heterogenous quantum network demanding fitting information transfer between the two encodings. Therefore, these hybrid techniques shed new light towards realizing scalable QIP and quantum communications.

Although the endeavor of combining DV toolboxes with CV frameworks has been started two decades ago in generating non-Gaussian states, it becomes only recently feasible to form hybrid optical entanglement between particle- and wave-like qubits in experiments [10-16]. The DV-CV hybridization turns out to enable potentially more powerful hybrid protocols. Among them include demonstrations of remote state preparation [14], quantum teleportation [12,13], entanglement swapping [16] and Einstein-Podolsky-Rosen (EPR) steering [15]. Despite of these impressive achievements, unfortunately in most of these demonstrations [8-12,14-22], the DVs are encoded in the superposition of a vacuum and single-photon number states and CVs are encoded in continuous quadrature components, except the work [13]. Alternatively, the success of these works is determined by the realization of single-photon entanglement. It is well known that single-photon detection cannot tell the difference between the vacuum state and single-photon state with detection loss. As such, such a DV entangled state in these hybrid schemes is incompatible with many existing QIP photonic platforms which make use of light polarizations or spatial/temporal modes for information encoding [1-4]. In addition, the involved DV and CV states can be by nature described within the same Hilbert space, owing to the fact that the vacuum and single-photon states can be expressed as CV wave functions in the phase representation and the CV states can be decomposed into the superposition of photon-number states in the Fock space. This hidden interconnection implies that the employed DV and CV spaces are not fully independent. Rather, they both rely on homodyne detection, thereby resulting in an unavoidable crosstalk in measurement, especially in the presence of photon loss and vacuum noise. Furthermore, the generation of hybrid DV-CV entanglement in these studies [8-12,14-22] are highly sensitive to phase fluctuations of two indistinguishable optical pathways before the beam splitter [17], which adversely make scalable QIP and quantum communications challenging.

We are aware of a recent work [13] that has been put forward with use of the DV qubits encoded in the polarizations. However, in this DV-CV scheme, the DV polarization modes are not in a true single photon state because they are from mixing of a weak coherent light beam and photons subtracted from a squeezed vacuum state. Therefore, its DV components are still incompatible with the existing true single photons based QIP photonic platforms [1-4].

Here we propose a new hybrid DV-CV scheme to bridge a single photon's discrete polarizations and continuous quadrature variables to breach these obstacles aforementioned. Unlike the previous proposals [10-12,14-22], we utilize two orthogonal polarizations of a single photon to encode the DV information. Different from the recent work with mixing weak coherent light [13], we create the hybrid entanglement using polarization entangled photon pairs and a joint two-photon Bell-state measurement (BSM). This choice helps to not only eliminate the effect of the vacuum and detection loss on measurement results, but also remove any possible crosstalk between two encoding spaces. Moreover, our DV-CV hybrid entanglement generation is based on two-photon Bell-state measurements (BSMs), which is not sensitive to optical path-length fluctuations. On the other hand, the CV encoding basis accordingly consists of even and odd Schr\"{o}dinger states with orthogonal polarizations. These advantages will enable our hybrid network to provide trustworthy links between single-photon-counting-based DV quantum computing nodes and CV quantum computing nodes. This new technique also opens up an alternative avenue for entanglement swapping, quantum state conversion and teleportation, promising a versatile platform for quantum networks.

\begin{figure*} [t]
\centering
\includegraphics[width=17 cm]{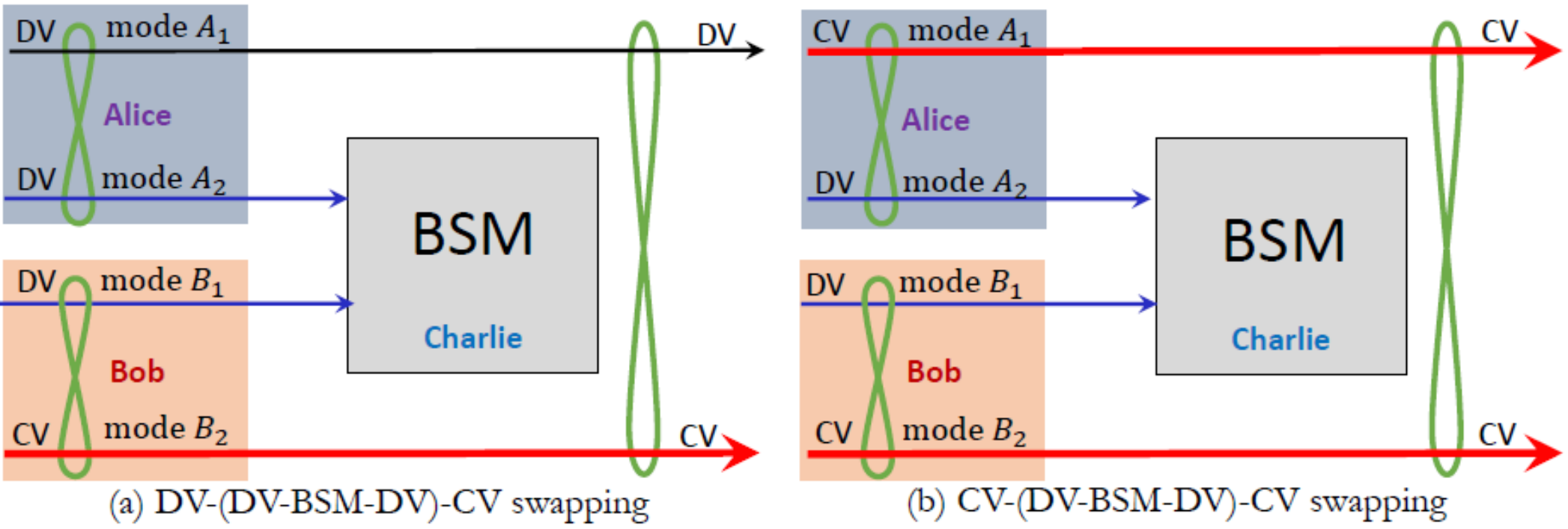}
\caption{(color online) Schematics of entanglement swapping assisted by hybrid DV-CV entanglement: (a) DV-(DV-BSM-DV)-CV swapping and (b) CV-(DV-BSM-DV)-CV swapping.}
\label{fig:figure2}
\end{figure*}

\emph{Hybrid entanglement generation.}  In our study, the ideal form of maximal hybrid entanglement as the target state is \begin{equation}\label{1}
|\Psi\rangle=\frac{1}{\sqrt{2}}(|1_H\rangle|\mathrm{Cat}^{+}_H\rangle+|1_V\rangle|\mathrm{Cat}^{-}_V\rangle),
\end{equation}
where $H/V$ represents the horizontal/vertical polarization, $\{|1_H\rangle, |1_V\rangle\}$ is the Fock basis for single-photon states, $|\mathrm{Cat}^{\pm}\rangle$ stand for positive and negative cat states $N^{\pm}(|\alpha\rangle\pm|-\alpha\rangle)$ with the normalization factors $N^{\pm}=1/\sqrt{2\pm 2e^{-2\alpha}}$ [23-25], and $\pm\alpha$ being amplitudes of coherent states $|\pm\alpha\rangle$, respectively.

To obtain the state (1), we are interested in the simple optical circuit illustrated in Fig. 1, where Alice (denoted by A hereafter) and Bob (B) are, respectively, using DV and CV encodings for QIP. In order to establish the internode link, they start with locally prepared nonclassical light states. Specifically, through spontaneous parametric down conversion (SPDC) [26] or spontaneous four-wave mixing (SFWM) [27], a two-photon polarization entangled state (or Bell state) $|\psi\rangle_{ab}=\frac{1}{\sqrt{2}}(|1_H\rangle_a|1_H\rangle_b+|1_V\rangle_a|1_V\rangle_b)$ is employed at Alice's side, where subscripts `$a$' and `$b$' indicate the spatial modes of paired photons; whereas a polarization-coupled Schr\"{o}dinger cat state $|\psi_B\rangle=\frac{1}{\sqrt{2}}(|\mathrm{Cat}^{+}_H\rangle+|\mathrm{Cat}^{-}_V\rangle)$ is adopted on Bob's site, which can be attained with a two-mode squeezed vacuum state [28] in an excellent approximation. A small fraction of light in $|\psi_B\rangle$ (on average one photon) is tapped off by a beam splitter (BS) with amplitude reflectivity $r\ll1$, and the state then evolves to $(1+r\hat{c}\hat{d}^{\dagger})|\psi_B\rangle_c|0\rangle_d$ with $\hat{c}$ and $\hat{d}^{\dagger}$ being the annihilation and creation operators for modes $c$ and $d$. Note that $\hat{c}|\mathrm{Cat}^{\pm}\rangle=\frac{\alpha N^{\mp}}{N^{\pm}}|\mathrm{Cat}^{\mp}\rangle$. On Alice's node, meantime, a polarization beam-splitter (PBS) is applied to direct the photon in mode $b$ into a center station where it is superposed with the subtracted photon in mode $d$ for a joint BSM. It is not difficult to show that in the center station, the two-photon coincidence events $C_{13}$ (or $C_{24}$) between single-photon detectors D$_1$ and D$_3$ (or D$_2$ and D$_4$) will project modes $a$ and $c$ onto the hybrid entangled state akin to Eq. (1),
\begin{equation}\label{2}
|\Psi^{(1)}\rangle_{ac}=\gamma_{+}|1_H\rangle_a|\mathrm{Cat}^{+}_H\rangle_c+\gamma_{-}|1_V\rangle_a|\mathrm{Cat}^{-}_V\rangle_c,
\end{equation}
except for the normalization. Here, $\gamma_{+}=r\alpha N^{+}/N^{-}$ and $\gamma_{-}=r\alpha N^{-}/N_{+}$. For $\alpha\gg1$, the above state begins to overlap with the form given in (1). The nonclassical and entanglement properties of this state can be experimentally approached with two single-photon detectors on Alice's node and two homodyne detectors on Bob's side. Thanks to the two-photon joint detection in the center station, it largely reduces the weight of the propagated vacuum state and makes photon losses less affect the negativity [29] of the state in its Wigner function representation. This coincidence counting measurement also ensures the negativity of the reconstructed density matrix after partial transposition to be close to the upper theoretical bound of 0.5 for the model state (2) when $\alpha\gg1$.

If placing a half-wave plate in mode $b$ before the first PBS, one can show that in this case the joint trigger clicks will herald the generation of
\begin{equation}\label{3}
|\Psi^{(3)}\rangle_{ac}=\gamma_{+}|1_H\rangle_a|\mathrm{Cat}^{+}_V\rangle_c+\gamma_{-}|1_V\rangle_a|\mathrm{Cat}^{-}_H\rangle_c.
\end{equation}
Similarly, another two hybrid entangled states
\begin{eqnarray}\label{4}
|\Psi^{(3)}\rangle_{ac}=\gamma_{-}|1_H\rangle_a|\mathrm{Cat}^{-}_H\rangle_c-\gamma_{+}|1_V\rangle_a|\mathrm{Cat}^{+}_V\rangle_c,\\
|\Psi^{(3)}\rangle_{ac}=\gamma_{+}|1_H\rangle_a|\mathrm{Cat}^{+}_V\rangle_c-\gamma_{-}|1_V\rangle_a|\mathrm{Cat}^{-}_H\rangle_c,
\end{eqnarray}
can be conditionally produced by preparing the Bob's initial state to be $|\psi'_B\rangle=\frac{1}{\sqrt{2}}(|\mathrm{Cat}^{+}_H\rangle-|\mathrm{Cat}^{-}_V\rangle)$ with the insertion of a half-wave plate in the channel $c$. Interestingly, the four states (2)-(5) form a complete set of DV-CV hybrid Bell entanglement. In addition, each of them resembles the spirit of the Schr\"{o}dinger Gedanken experiment on entangling two macroscopic states with a microscopic degree of freedom. Unless otherwise specified, the following discussions are mainly based on expression (2) by assuming $\gamma_{\pm}=1/\sqrt{2}$.

\emph{Hybrid entanglement swapping.}  Entanglement swapping [30-33], at the heart of quantum repeaters [34,35], aims to transfer entanglement to two parties that are produced independently and never interacted previously. In the course of a hybrid network, the swapping technique is the key to link and transfer entanglement between diverse nodes.

Dissimilar to the past demonstrations [16,33,36], our hybrid entanglement swapping is accomplished by interfering one DV port from each bipartite system subsequently followed a two-photon BSM. Dependent on the Bell states prepared on the nodes of Alice and Bob, we propose two different solutions to realize hybrid entanglement swapping (see Fig. 2). In the first solution, as schematic in Fig. 2(a), Alice and Bob first locally prepare their respective Bell pairs, say, $|\psi\rangle_{A_1A_2}$ and $|\Psi^{(1)}\rangle_{B_1B_2}$. The resulting initial state is $|\Phi\rangle_{in}=|\psi\rangle_{A_1A_2}|\Psi^{(1)}\rangle_{B_1B_2}$. Next, they send the DV qubits $A_2$ and $B_1$ to a third party Charlie. Charlie then carries out a Bell projection on the two receiving qubits. By chance the measurement outcome will swap the original entanglements to the two remaining modes $A_1$ and $B_2$, i.e., $\langle\psi|_{A_2B_1}|\Phi\rangle_{in}=|\Psi^{(1)}\rangle_{A_1B_2}$. As one can see, the two end nodes eventually arrive at sharing the DV-CV hybrid entanglement jointly in spite of one of them beginning with a DV-only system. The conditioned state $|\Psi^{(1)}\rangle_{A_1B_2}$ may be experimentally verified and characterized by quantum state tomography [37] in the relevant basis (single-photon polarizations $\{|1_H\rangle,|1_V\rangle\}$ for DV and coherent-state superpositions $\{|\mathrm{Cat}^{+}_H\rangle,|\mathrm{Cat}^{-}_V\rangle\}$ for CV). In the case of the other three BSM outcomes, Charlie could communicate the measurement results with Alice and Bob, and suggest them to make local corrections by performing proper Pauli operations. Due to the nature of the initially involved states, we call this scheme as the DV-(DV-BSM-DV)-CV swapping. In the second scenario as sketched in Fig. 2b, Alice and Bob both start with the same hybrid entangled states $|\Psi^{(1)}\rangle_{A_1A_2}$ and $|\Psi^{(1)}\rangle_{B_1B_2}$, and route their DV parts $A_2$ and $B_1$ to Charlie for a two-photon BSM. Following the similar procedure, one can show that an entangled cat state can be established to link the two CV end nodes even far way from each other, $|\psi\rangle_{A_1B_2}=(|\mathrm{Cat}^{+}_H\rangle_{A_1}|\mathrm{Cat}^{+}_H\rangle_{B_2}+|\mathrm{Cat}^{-}_V\rangle_{A_1}
|\mathrm{Cat}^{-}_V\rangle_{B_2})/\sqrt{2}$, provided by the successful Bell measurement on Charlie's station. Such entanglement can be revealed by fulfilling quantum state tomography in the basis of $\{|\mathrm{Cat}^{+}_H\rangle,|\mathrm{Cat}^{-}_V\rangle\}$ on each CV node. In the light of the first protocol, we accordingly name this one as the CV-(DV-BSM-DV)-CV swapping. Although the hybrid entanglement (2) is picked to lay out the two swapping schemes, one can examine that the algorithms also work for other hybrid Bell triplet states (3)-(5).

\emph{Hybrid quantum teleportation.}  Quantum teleportation [38-42] is able to transfer quantum information between different physical systems via shared entanglement and classical communication channels. Though it is central to many quantum technologies including quantum communications and QIP, the most realizations [42] based on either CV or DV solution have their own strengths and limitations. By taking advantages of both, hybrid quantum techniques [13,18,22,41] offer a versatile and compelling way to mitigate their downsides. As such, hybrid quantum teleportation [42] has become superior to the mainstream protocols. Here, we develop a hybrid teleportation technique, akin to the recent demonstration [13], which is compatible to the current platforms of QIP and quantum communications.

Our protocol is built upon one of the auxiliary states (2)-(5). Let us suppose a single-photon polarization qubit in Alice's possession that will be conveyed to Bob,
\begin{equation}\label{6}
|\psi\rangle_A=c_H|1_H\rangle_{A}+c_V|1_V\rangle_A,
\end{equation}
where the complex numbers $c_H$ and $c_V$ satisfying the normalization condition $|c_H|^2+|c_V|^2=1$ are unknown to either of them. Their locations will be entangled through any one of the resource states (2)-(5), which is fixed in advance by the mutual agreement. To ease the discussions, let us assume that $|\Psi^{(1)}\rangle_{CB}$ will be shared by Alice and Bob. In particular, the DV part is transmitted to Alice while the CV part is retained for Bob. Now, the state of the total system becomes $|\psi\rangle_A|\Psi^{(1)}\rangle_{CB}=(c_H|1_H\rangle_{A}+c_V|1_V\rangle_A)\bigotimes\frac{1}{\sqrt{2}}(|1_H\rangle_{C}
|\mathrm{Cat}^{+}_{H}\rangle_{B}+|1_V\rangle_{C}|\mathrm{Cat}^{-}_{V}\rangle_{B})$.
To perform the teleportation, Alice will then carry out a local measurement on two available single photons in the DV Bell basis defined by
\begin{eqnarray}\label{7}
|\Omega^{\pm}\rangle_{AC}&=&(|1_H\rangle_A|1_H\rangle_{C}\pm|1_V\rangle_A|1_V\rangle_{C})/\sqrt{2},\\ |\Theta^{\pm}\rangle_{AC}&=&(|1_H\rangle_A|1_V\rangle_{C}\pm|1_V\rangle_A|1_H\rangle_{C})/\sqrt{2}.
\end{eqnarray}
By recasting the state of Alice's two quibits in terms of the superpositions of Bell basis, the total state assumes the following form: $|\psi\rangle_A|\Psi^{(1)}\rangle_{CB}=[|\Omega^+\rangle_{AC}(c_H|\mathrm{Cat}^+_H\rangle_B+c_V|\mathrm{Cat}^-_V\rangle_B)+
|\Omega^-\rangle_{AC}(c_H|\mathrm{Cat}^+_H\rangle_B-c_V|\mathrm{Cat}^-_V\rangle_B)+|\Theta^+\rangle_{AC}(c_H|\mathrm{Cat}^-_V\rangle_B
+c_V|\mathrm{Cat}^+_H\rangle_B)+|\Theta^-\rangle_{AC}(c_H|\mathrm{Cat}^-_V\rangle_B-c_V|\mathrm{Cat}^+_H\rangle_B)]/2$. The actual teleportation takes place when Alice accomplishes her BSM on $A$ and $C$ in the four Bell basis. Apparently, Alice's local detection induces the above total state to collapse to one of the four components with equal probability, and enforces the state of the Bob's CV qubit to pick up the coefficients of the DV state (6). Given the above expression, there are four possible states for Bob's qubit, corresponding to transformation images of the source state. By communicating with Alice on her result through a classical channel, Bob will know which of the four states that his qubit belongs to and then accordingly, implement a unitary operation to recover it to the desired state $|\psi\rangle_B=c_H|\mathrm{Cat}^+_H\rangle_B+c_V|\mathrm{Cat}^-_V\rangle_B$. To this end, the teleportation is completed.

\begin{figure} [t]
\centering
\includegraphics[width=8.5 cm]{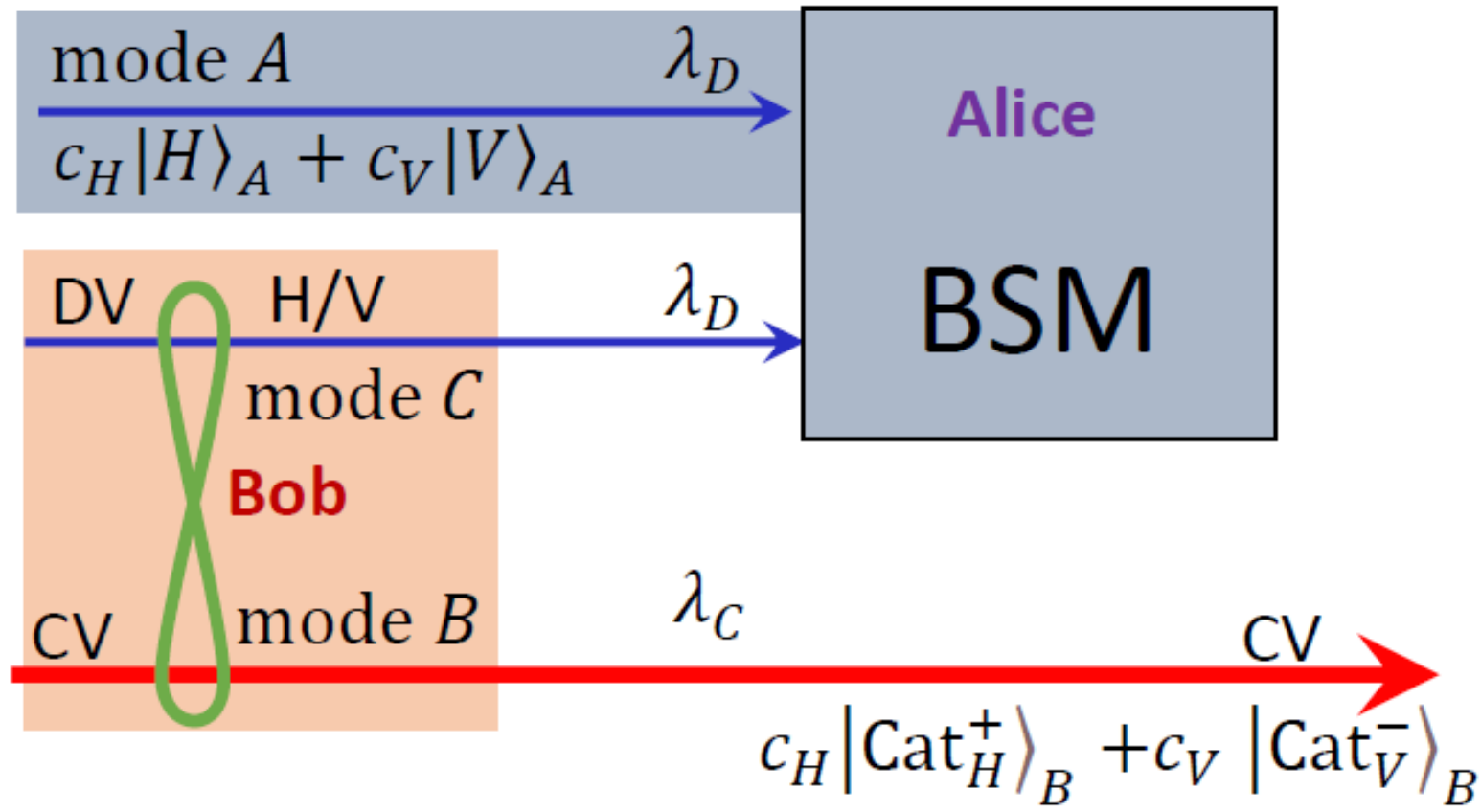}
\caption{(color online) Schematic of teleporting an unknown single-photon polarization qubit via hybrid DV-CV quantum entanglement.}
\label{fig:fig3}
\end{figure}


We notice that with some modifications, the teleportation scheme can be geared towards applications for quantum wavelength conversion between a single-photon polarization qubit and a two-mode squeezing with use of a CV quantum teleportor. For a theoretical analysis, please refer to Ref. [43] for details.

In summary, we have shown how to hybridize an optical polarization qubit and a CV cat qubit as a new entanglement resource based on the herald of two-photon Bell polarization-state measurements. Compared with most of the previous protocols based on the single-photon entanglement, our solution is not only compatible with the existing frameworks of QIP and communication network, but also enables a flexible hybrid technological development. In this regard, we especially considered entanglement swapping and quantum teleportation as examples. Nevertheless, our proposal could serve as an attractive alternative for the realization of heterogenous quantum network empowered by intriguing properties of hybrid DV-CV entanglement but avoiding shortcomings of each individual.

J.W. was supported by NSF (1806519 and EFMA-1741693). I.N. acknowledges financial support from AFOSR(FA9550-19-1-0066). C.Q. was supported by NSF (CNS-1750704 and CPS-1932447). C.Z. was supported by AFOSR (FA9550-20-1-0220), NSF (PHY-1806227), and ARO (W911NF-17-1-0128). S.D. acknowledges the STARs (40071050) and Start-Up (40037055) fundings from The University of Texas at Dallas.


\begin{thebibliography}{99}

\bibitem{1}   M. A. Nielsen and I. L. Chuang, \textit{Quantum Computation and Quantum Information} (Cambridge University Press, 2010).
\bibitem{2}   P. Kok, W. J. Munro, K. Nemoto, T. C. Ralph, J. P. Dowling, and G. J. Milburn, ``Linear optical quantum computing with photonic qubits," Rev. Mod. Phys. \textbf{79}, 135 (2007).
\bibitem{3}   H. J. Kimble, ``The quantum internet," Nature \textbf{453}, 1023 (2008).
\bibitem{4}   J. L. O'Brien, A. Furusawa, and J. Vuckovi\'{c}, ``Photonic quantum technologies," Nat. Photonics \textbf{3}, 687 (2009).
\bibitem{5}   S. L. Braunstein and P. van Loock, ``Quantum information with continuous variables," Rev. Mod. Phys. \textbf{77}, 513 (2005).
\bibitem{6}   C. Weedbrook, S. Pirandola, R. Garc\'{i}a-Patr\'{o}n, N. J. Cerf, T. C. Ralph, J. H. Shapiro, and S. Lloyd, ``Gaussian quantum information," Rev. Mod. Phys. \textbf{84}, 621 (2021).
\bibitem{7}   U. L. Andersen, G. Leuchs, and C. Silberhorn, ``Continuous-variable quantum information processing," Laser Photon. Rev. \textbf{4}, 337 (2010).
\bibitem{8}   U. L. Andersen, J. S. Neergaard-Nielsen, P. van Loock, and A. Furusawa, ``Hybrid discrete- and continuous-variable quantum information," Nat. Phys. \textbf{11}, 713 (2015).
\bibitem{9}   P. van Loock, ``Optical hybrid approaches to quantum information," Laser Photon. Rev. \textbf{5}, 167 (2011).
\bibitem{10}  H. Jeong, A. Zavatta, M. Kang, S.-W. Lee, L. S. Costanzo, S. Grandi, T. C. Ralph, and M. Bellini, ``Generation of hybrid entanglement of light," Nat. Photonics \text{8}, 564 (2014).
\bibitem{11}  O. Morin, K. Huang, J. Liu, H. L. Jeannic, C. Fabre, and J. Laurat, ``Remote creation of hybrid entanglement between particle-like and wave-like optical qubits," Nat. Photonics \textbf{8}, 570 (2014).
\bibitem{12}  A. E. Ulanov, D. Sychev, A. A. Pushkina, I. A. Fedorov, and A. I. Lvovsky, ``Quantum teleportation between discrete and continuous encodings of an optical qubit," Phys. Rev. Lett. \textbf{118}, 160501 (2017).
\bibitem{13}  D. V. Sychev, A. E. Ulanov, E. S. Tiunov, A. A. Pushkina, A. Kuzhamuratov, V. Novikov, and A. I. Lvovsky, "Entanglement and teleportation between polarization and wave-like encondings of an optical qubit," Nat. Commun. \textbf{9}, 3672 (2018).
\bibitem{14}  H. Le Jeannic, A. Cavaill\`{e}s, J. Raskop, K. Huang, and J. Laurat, ``Remote preparation of arbitrary continuous-variable qubits using loss-tolerant hybrid entanglement of light," Optica \textbf{5}, 1012 (2018).
\bibitem{15}  A. Cavaill\`{e}s, H. Le Jeannic, J. Raskop, G. Guccione, D. Markham, E. Diamanti, M. D. Shaw, V. B. Verma, S. W. Nam, and J. Laurat, ``Demonstration of Einstein-Podolsky-Rosen steering using hybrid continuous- and discrete-variable entanglement of light," Phys. Rev. Lett. \textbf{121}, 170403 (2018).
\bibitem{16}  G. Guccione, T. Darras, H. Le Jeannic, V. B. Verma, S. W. Nam, A. Cavaill\`{e}s, and J. Laurat, ``Connecting heterogenous quantum networks by hybrid entanglement swapping," Sci. Adv. \textbf{6}, eaba4508 (2020).
\bibitem{17}  K. Huang, H. Le Jeannic, O. Morin, T. Darras, G. Guccione, A. Cavaill\`{e}s, and J. Laurat, ``Engineering optical hybrid entanglement between discrete- and continuous-variable states," New J. Phys. \textbf{21}, 083033 (2019).
\bibitem{18}  S.-W. Lee and H. Jeong, ``Near-deterministic quantum teleportation and resource-efficient quantum computation using linear optics and hybrid qubits," Phys. Rev. A \textbf{87}, 022326 (2013).
\bibitem{19}  K. Kries and P. van Loock, ``Classifying, quantifying, and witnessing qudit-qumode hybrid entanglement," Phys. Rev. A \textbf{85}, 032307 (2012).
\bibitem{20} U. L. Andersen and J. S. Neergaard-Nielsen, ``Heraded generation of a micro-macro entangled state," Phys. Rev. A \textbf{88}, 022337 (2013).
\bibitem{21}   L. S. Costanzo, A. Zavatta, S. Grandi, M. Bellini, H. Jeong, M. Kang, S.-W. Lee, and T. C. Ralph, "Properties of hybrid entanglement between discrete- and continuous-variable states of light," Phys. Scr. \textbf{90}, 074045 (2015).
\bibitem{22}  K. Park, S.-W. Lee, and H. Jeong, "Quantum teleportation between particlelike and fieldlike qubits using hybrid entanglement under decoherence effects," Phys. Rev. A \textbf{86}, 062301 (2012).
\bibitem{23}  A. Ourjoumtsev, R. Tualle-Brouri, J. Laurat, and P. Grangier, ``Generating optical Schr\"{o}dinger kittens for quanutm information processing," Science \textbf{312}, 83 (2006).
\bibitem{24}  J. S. Neergaard-Nielsen, B. M. Nielsen, C. Hettich, K. M\`{o}lmer, and E. S. Polzik, ``Generation of a superposition of odd photon number states for quantum information networks," Phys. Rev. Lett. \textbf{97}, 083604 (2006).
\bibitem{25}  B. Hacker, S. Welte, S. Daiss, A. Shaukat, S. Ritter, L. Li, and G. Rempe, ``Deterministic creation of entangled atom-light Schr\"{o}dinger-cat states," Nat. Photonics \textbf{13}, 110 (2019).
\bibitem{26}  P. G. Kwiat, K. Mattle, H. Weinfurter, A. Zeilinger, A. V. Sergienko, and Y. Shih, ``New high-intensity source of polarization-entangled photon pairs," Phys. Rev. Lett. \textbf{75}, 4337 (1995).
\bibitem{27}  S. Du, J. Wen, and M. H. Rubin, ``Narrowband biphoton generation near atomic resonance," J. Opt. Soc. Am. B \textbf{25}, C98 (2008).
\bibitem{28}  M. Zhang, M. A. Guidry, R. N. Lanning, Z. Xiao, J. P. Dowling, I. Novikova, and E. E. Mikhailov, ``Multi-pass configuration for improved squeezed vacuum genertion in hot Rb vapor," Phys. Rev. A \textbf{96}, 013835 (2017).
\bibitem{29}  G. Vidal and R. F. Werner, ``Computable measure of entanglement," Phys. Rev. A \textbf{65}, 032314 (2002).
\bibitem{30}  M. \.{Z}ukowski, Z. Zeilinger, M. A. Horne, and A. K. Ekert, ````Event-ready-detectors" Bell experiment via entanglement swapping," Phys. Rev. Lett. \textbf{71}, 4287 (1993).
\bibitem{31}  J.-W. Pan, D. Bouwmeester, H. Weinfurter, and A. Zeilinger, ``Experimental entanglement swapping: Entangling photons that never interacted," Phys. Rev. Lett. \textbf{80}, 3891 (1998).
\bibitem{32}  X. Jia, X. Su, Q. Pan, J. Gao, C. Xia, and K. Peng, ``Experimental demonstration of unconditional entanglement swapping for continuous variables," Phys. Rev. Lett. \textbf{93}, 250503 (2004).
\bibitem{33}  S. Takeda, M. Fuwa, P. van Loock, and A. Furusawa, ``Entanglement sawpping between discrete and continuous variables," Phys. Rev. Lett. \textbf{114}, 100501 (2015).
\bibitem{34}  H.-J. Briegel, W. D\"{u}r, J. I. Cirac, and P. Zoller, ``Quantum repeaters: The tole of imperfect local operations in quantum communication," Phys. Rev. Lett. \textbf{81}, 5932 (1998).
\bibitem{35}  S. Wehner, D. Elkouss, and R. Hanson, ``Quantum internet: A vision for the road ahead," Science \textbf{362}, eaam9228 (2018).
\bibitem{36}  Y. Lim, J. Joo, T. P. Spiller, and H. Jeong, ``Loss-resilient photonic entanglement swapping using optical hybrid states," Phys. Rev. A \textbf{94}, 062337 (2016).
\bibitem{37}  A. I. Lvovsky and M. G. Raymer, ``Continuous-variable optical quantum state tomography," Rev. Mod. Phys. \textbf{81}, 299 (2009).
\bibitem{38}  C. H. Bennett, G. Brassard, C. Cr\'{e}peau, R. Jozsa, A. Peres, and W. K. Wooters, ``Teleporting an unknown quantum state via dual classical and Einstein-Podolsky-Rosen channels," Phys. Rev. Lett. \textbf{70}, 1895 (1993).
\bibitem{39}  D. Boschi, S. Branca, F. De Martini, L. Hardy, and S. Popescu, ``Experimental realization of teleporting an unknown pure quantum state via dual classical and Einstein-Podolsky-Rosen channels," Phys. Rev. Lett. \textbf{80}, 1121 (1998).
\bibitem{40}  D. Bouwmeester, J.-W. Pan, K. Mattle, M. Eibl, H. Weinfurter, and A. Zeilinger, ``Experimental quantum teleportation," Nature \textbf{390}, 575 (1997).
\bibitem{41}  S. Takeda, T. Mizuta, M. Fuwa, P. van Loock, and A. Furusawa, ``Deterministic quantum teleportation of photonic quantum bits by a hybrid technique," Nature \textbf{500}, 315 (2013).
\bibitem{42}  S. Pirandola, J. Eisert, C. Weedbrook, A. Furusawa, and S. L. Braunstein, ``Advances in quantum teleportation," Nat. Photonics \textbf{9}, 641 (2015).
\bibitem{43}  X.-W. Luo, C. Zhang, I. Novikova, C. Qian, and S. Du, ``Wavelength conversion for single-photon polarization qubits through continuous variable quantum teleportation" (unpublished).

\end{thebibliography}
\end{document}